\documentclass[aip, jcp, amsmath,amssymb,reprint]{revtex4-1}

\draft 
\usepackage{graphicx}
\usepackage{dcolumn}
\usepackage{bm}
\usepackage{hyperref}
\usepackage[normalem]{ulem}

\usepackage[utf8]{inputenc}
\usepackage[T1]{fontenc}
\usepackage{mathptmx}
\usepackage[english]{babel}
\usepackage{xcolor}

\begin{document}


\title{Phase Behaviour of Binary Hard-Sphere Mixtures: Free Volume Theory Including Reservoir Hard-Core Interactions} 



\author{J. Opdam}
\affiliation{ 
Laboratory of Physical Chemistry, Department of Chemical Engineering and Chemistry, \& Institute for Complex Molecular Systems (ICMS), Eindhoven University of Technology, P.O. Box 513, 5600 MB, Eindhoven, The Netherlands
}

\author{M. P. M. Schelling}
\affiliation{ 
Laboratory of Physical Chemistry, Department of Chemical Engineering and Chemistry, \& Institute for Complex Molecular Systems (ICMS), Eindhoven University of Technology, P.O. Box 513, 5600 MB, Eindhoven, The Netherlands
}

\author{R. Tuinier}
\email{r.tuinier@tue.nl}
\affiliation{ 
Laboratory of Physical Chemistry, Department of Chemical Engineering and Chemistry, \& Institute for Complex Molecular Systems (ICMS), Eindhoven University of Technology, P.O. Box 513, 5600 MB, Eindhoven, The Netherlands
}


\date{\today}

\begin{abstract}
Comprehensive calculations were performed to predict the phase behaviour of large spherical colloids mixed with small spherical colloids that act as depletant. To this end, the free volume theory (FVT) of Lekkerkerker \textit{et al.} [\textit{Europhys. Lett.} \textbf{20} (1992) 559] is used as a basis and is extended to explicitly include the hard-sphere character of colloidal depletants into the expression for the free volume fraction. Taking the excluded volume of the depletants into account in $both$ the system and the reservoir provides a relation between the depletant concentration in the reservoir and in the system that accurately matches with computer simulation results of Dijkstra \textit{et al.} [\textit{Phys. Rev. E} \textbf{59} (1999) 5744]. Moreover, the phase diagrams for highly asymmetric mixtures with size ratios $q \lesssim 0.2$ obtained by using this new approach corroborates simulation results significantly better than earlier FVT applications to binary hard-sphere mixtures. The phase diagram of a binary hard-sphere mixture with a size ratio of $q = 0.4$, where a binary interstitial solid solution is formed at high densities, is investigated using a numerical free volume approach. At this size ratio, the obtained phase diagram is qualitatively different from previous FVT approaches for hard-sphere and penetrable depletants, but again compares well with simulation predictions. 
\end{abstract}


\maketitle 

\section{Introduction}
\noindent Colloidal particles are ubiquitously present in everyday products such as cosmetics, foodstuffs and coatings\cite{Piazza2011}. The stability of these products depends on the phase behaviour of the colloids that can undergo phase transitions similar to atomic or molecular systems\cite{Poon2004a}. The conditions where these transitions occur are determined by the (effective) interactions between the colloidal particles and are affected by their environment \cite{Lekkerkerker2011,GonzalezGarcia2016,Tamura2020}. In colloidal products this environment is often composed of different types of other colloidal particles that interact with each other. For example, coating formulations often contain multiple colloidal components that can serve either as binder, pigment or as additive \cite{deWith2018}. These colloidal particles can be spherical but are also often anisotropic in the case of pigments. Moreover, the size ratio between the different particles is often quite large, for example for stratification purposes \cite{Cardinal2010,Schulz2018}. In these types of colloidal mixtures, depletion interactions are present that can lead to phase separation which is often undesired in colloidal products since it leads to inhomogeneities. However, depletion interactions can also be used advantageously, for example in the separation or fractionation of colloidal mixtures\cite{Bibette1991,Park2010} or inducing protein crystallization\cite{Tanaka2002}. For the optimal use of colloidal mixtures it is crucial to have a good understanding of the phase behaviour of the colloidal particles.

\begin{figure}[tb!]
	\centering
		\includegraphics[width=.48\textwidth]{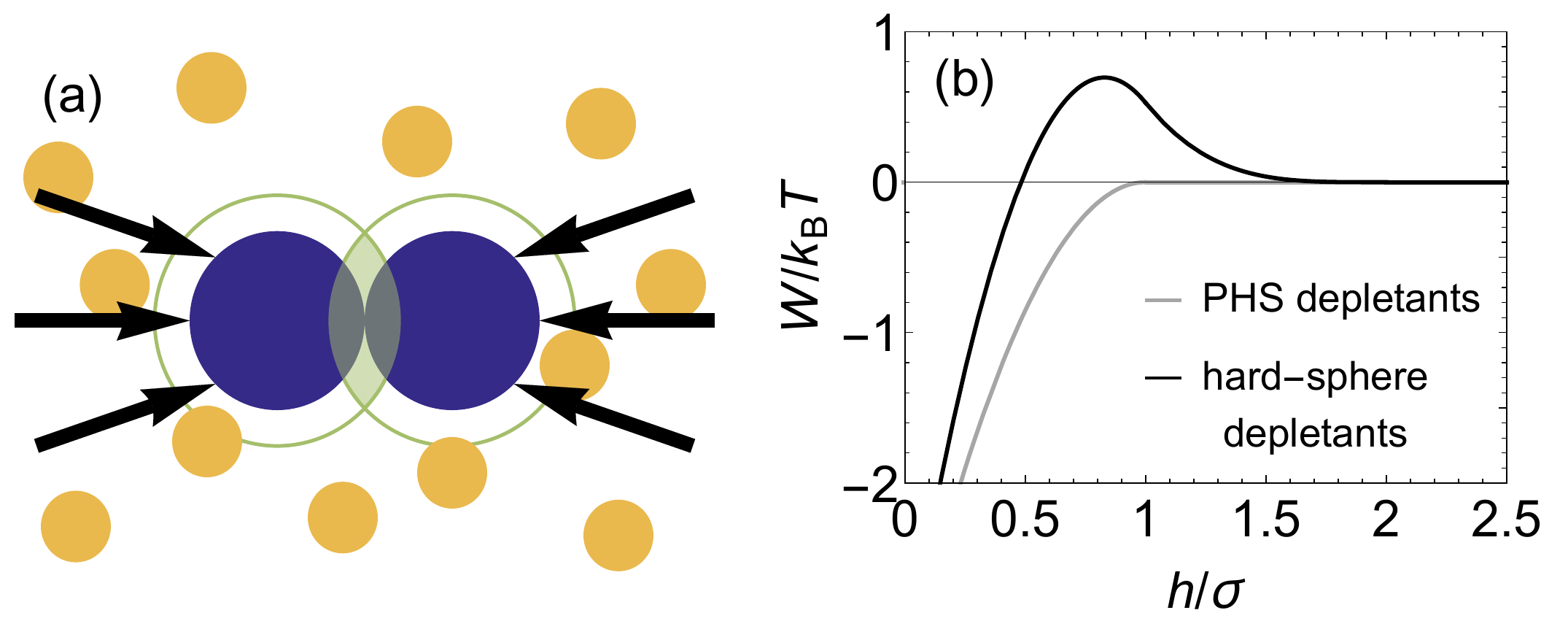}
	\caption{(a) Schematic representation of the depletion interaction mediated by hard-sphere or penetrable hard sphere depletants. The depletion zones around the large spheres that are inaccessible for the small spheres are shown by the thick green circles and the overlap of the depletion zones resulting in more free volume available for the small spheres is highlighted. (b) Comparison between the Asakura--Oosawa \cite{Asakura1954} depletion potential for PHS depletants (gray curve) and the depletion potential induced by hard-sphere depletants obtained from a second order perturbation theory (black curve) \cite{Mao1995} for a size ratio of $q = 0.1$ and depletant volume fraction of $\pi/15$. The distance between the colloidal particles $h$ is normalized by the diameter of the depletants $\sigma$.}
	\label{fig1}
\end{figure}

Colloidal particles have a finite volume and therefore excluded volume interactions are always present in colloidal systems. Excluded volume is generally associated to repulsive interactions, but in mixtures of colloids with different sizes excluded volume interactions can also indirectly induce effective attractions between particles. Around the larger colloidal particles a depletion zone exists that is inaccessible to the centers of the smaller particles. Once the depletion zones of different particles overlap, the total volume available for the smaller particles will increase leading to an effective depletion attraction\cite{Asakura1954,Asakura1958,Vrij1976,Lekkerkerker2011} between the larger colloids, induced by the excluded volume repulsion between the large and small particles. The depletion interaction is schematically illustrated in Fig. \ref{fig1}a. Due to the depletion attraction, colloidal particles can undergo phase transitions at much lower concentrations then expected for the single component dispersion. 

In 1954 Asakura and Oosawa\cite{Asakura1954} were first to theoretically consider the interaction between two spherical particles as mediated by non-adsorbing macromolecules and showed this leads to an effective attraction due to excluded volume interactions. A few years later\cite{Asakura1958} they explicitly quantified the case of the interaction between two hard spheres mediated by other hard spheres in the dilute limit. In their pioneering paper, Asakura and Oosawa speculated already about more complicated situations. Here we consider in some detail the phase behaviour of Asymmetric binary hard sphere dispersions. 

An insightful and relatively simple method to elucidate the phase behaviour of mixtures with colloids and depletants is free volume theory (FVT) \cite{Lekkerkerker1992}. In FVT, phase behaviour is determined using a thermodynamic description of the system containing colloids and depletants, based on estimating the free volume available for the depletants. FVT was originally developed to study colloid-polymer mixtures, where the colloids were described as hard spheres and the polymers were considered as penetrable hard spheres\cite{Lekkerkerker1992}. Penetrable hard spheres (PHS) are defined as non-additive spheres that can freely overlap with each other but cannot overlap with the colloidal particles\cite{Vrij1976}, an approximation that is reasonably accurate if the polymers are ideal chains or are dilute. The PHS model was later applied in FVT studies of mixtures with polymers that have a large size compared to the colloidal particles\cite{Moncho-Jorda2003} and colloidal particles that have additional interactions besides the excluded volume interactions\cite{GonzalezGarcia2016}. FVT has also been extended towards using interacting polymers as depletants\cite{Fleer2007,Fleer2008} and FVT approaches to describe the phase behaviour of binary colloidal hard-sphere mixtures have been proposed \cite{Lekkerkerker1993,Poon1994,Lekkerkerker2002}. Although the nature of the depletion interaction is similar for PHS and hard-sphere depletants, the inclusion of excluded volume of the depletants has a significant effect on the free volume available to the depletants. Moreover, the excluded volume repulsion between the depletants leads to a significant change in the effective pair potential between two hard spheres mixed with depletants \cite{Mao1995,Biben1996,Gotzelmann1998,Roth2000}, as shown in Fig. \ref{fig1}b. The range of the depletion attraction for hard-sphere depletants is significantly smaller and a repulsive barrier is present in the pair potential. Computer simulations on binary hard-sphere mixtures revealed that there is not only a primary minimum and a primary maximum in the pair potential, but more concentration-dependent oscillations around a zero potential are present\cite{Biben1996}. 

The phase behaviour of binary hard-sphere mixtures has been widely studied as a fundamental problem. For a long time it was believed that binary hard-sphere mixtures are thermodynamically stable for all concentrations and size ratios \cite{Lebowitz1964}, until Biben and Hansen\cite{Biben1990,Biben1991} first showed that phase separation can occur in binary hard-sphere mixtures. This finding was later confirmed by a variety of theoretical approaches\cite{Velasco1999, Roth2001, Suematsu2016a}, simulation studies\cite{Biben1996,Dijkstra1998,Dijkstra1999} and experimental works\cite{Duijneveldt1993,Imhof1995}. A historical overview of studies on binary hard-sphere mixtures can be found in \cite{Dijkstra1999}. Most of these studies have been focused on highly asymmetric binary hard-sphere mixtures with size ratios $q\lesssim 0.2$ where pair potential based methods can still be applied, as shown by Dijkstra \textit{et al.} \cite{Dijkstra1999}. An exact derivation of the AO potential was done in the canonical ensemble\cite{Rovigatti2015a} and in the semi-grand canonical ensemble\cite{Dijkstra1999}. For larger size ratios the assumption of pairwise additivity becomes less accurate due to the possibility of overlap of multiple depletion zones leading to many-body interactions\cite{Meijer1994}. Even for hard-sphere + PHS mixtures pairwise additivity\cite{Dijkstra1999b} of the interaction is only exact for size ratios $q<0.154$. FVT does not rely on an (effective) pair potential for the large spheres, but the depletants are explicitly incorporated through a thermodynamic description of the binary system and multi-body interactions are taken into account. Moreover, it has been shown that anisotropy in particle shape can be taken into account in a relatively simple manner with FVT \cite{Vliegenthart1999,Oversteegen2005,Garcia2018,GonzalezGarcia2018}. For these reasons, FVT is a promising and versatile method to gain insight in the phase behaviour of colloidal mixtures. However, already for the binary hard-sphere mixture there is a significant discrepancy between the FVT results in comparison with results from simulations\cite{Dijkstra1999} and perturbation theory\cite{Velasco1999}. 

In this paper, we first show why the original FVT approach to account for the excluded volume of colloidal depletants\cite{Lekkerkerker1993,Poon1994,Lekkerkerker2002} does not lead to an accurate description of the phase behaviour for binary hard-sphere mixtures. Next, we propose a FVT approach that accurately takes the excluded volume of the depletants into account. We focus on highly asymmetric binary hard-sphere mixtures ($q\lesssim 0.2$) for comparison with previous studies to validate the proposed approach. We also briefly discuss the possibility of applying FVT for a larger size ratio of 0.4 where an interstitial solid solution is formed at high densities \cite{Filion2009,Filion2011}. 

\section{Theory}\label{sec:theory}
\noindent In this section we provide an overview of the FVT used in this paper. In FVT the system of interest containing colloidal particles and depletants is assumed to be in thermodynamic equilibrium with a hypothetical reservoir through a membrane that is permeable to the solvent and depletants, but impermeable to the large colloidal particles. This equilibrium is used as a starting point for the derivation of the thermodynamic properties of the binary system. First, we provide the original equations of FVT for hard spheres mixed with penetrable hard spheres \cite{Lekkerkerker1992,Lekkerkerker2011} and an adjusted description of the free volume in a face-centered-cubic (FCC) crystal based on geometrical arguments \cite{Garcia2018a}. Then we show the correction on the semi-grand potential for hard-sphere depletants first discussed by Lekkerkerker and Stroobants \cite{Lekkerkerker1993} and argue why this correction is not sufficient to accurately describe binary hard-sphere mixtures. Finally, we provide a novel description of the semi-grand potential and free volume fraction in a binary hard-sphere system and explain how this can be used to calculate phase coexistence binodals. The focus in this paper is on highly asymmetric binary hard-sphere mixtures ($q \lesssim 0.2)$, however, we also briefly discuss the applicability of FVT for a larger size ratio of $q = 0.4$. All calculations were performed using Wolfram Mathematica 12.

\subsection{Semi-grand potential}
\noindent The semi-grand potential $\Omega$ describing a system containing $N_{\mathrm{c}}$ colloidal particles and $N_{\mathrm{d}}$ depletants, in contact with a depletant reservoir, is a Legendre transformation of the Helmholtz free energy $F$:

\begin{equation}
\Omega(N_{\mathrm{c}}, V, T, \mu_{\mathrm{d}}) = F(N_{\mathrm{c}}, N_{\mathrm{d}}, V, T) - \mu_{\mathrm{d}} N_{\mathrm{d}} 
\quad\text{,}
\label{semigrandpot1} 
\end{equation}

\noindent where $\mu_{\mathrm{d}}$ denotes the chemical potential of depletants in the system, and the volume and temperature are given by $V$ and $T$, respectively. In this approach, the solvent is treated as background. From Eq. \ref{semigrandpot1}, the following thermodynamic relation is obtained:

\begin{equation}
\left( \frac{\partial \Omega}{\partial \mu_{\mathrm{d}}} \right)_{N_{\mathrm{c}}, V, T} = - N_{\mathrm{d}}
\quad\text{,}
\end{equation}

\noindent from which it follows that:

\begin{equation}
\Omega(N_{\mathrm{c}}, V, T, \mu_{\mathrm{d}}) = F_{0}(N_{\mathrm{c}}, V, T) - \int\displaylimits_{-\infty}^{\mu_{\mathrm{d}}} N_{\mathrm{d}}(\mu'_{\mathrm{d}}) \, \mathrm{d} \mu'_{\mathrm{d}}
\quad\text{,}
\label{semigrandpot2}
\end{equation}

\noindent where the equality $\Omega(N_{\mathrm{c}}, V, T, \mu_{\mathrm{d}} \rightarrow{} -\infty) = F_{0}(N_{\mathrm{c}}, V, T)$ was used. In this equation, $F_{0}$ is the Helmholtz free energy of a pure system of hard spheres (i.e. colloids without depletants). Eqs. \ref{semigrandpot1}-\ref{semigrandpot2} are exact and hold for any type of binary colloidal mixture.

\subsubsection{Penetrable hard spheres} \label{sssec:PHS}
\noindent In original FVT \cite{Lekkerkerker1992}, which was developed to describe mixtures of colloids and polymers, polymers were described as penetrable hard spheres (PHS) that can freely overlap with each other but cannot overlap with the colloidal particles. To obtain an expression for the number of depletants in the system ($N_{\mathrm{d}}$), Widom's insertion theorem \cite{Widom1963} is used, which gives for the chemical potential of PHS depletants in the system:

\begin{equation}
\mu_{\mathrm{d}} = \mathrm{const} + k_{\mathrm{B}}T \, \mathrm{ln} \frac{N_{\mathrm{d}}}{\langle V_{\mathrm{free}} \rangle}
\label{chemPotentialSystem}
\quad\text{,}
\end{equation}

\noindent where $\langle V_{\mathrm{free}} \rangle$ is the ensemble-averaged volume that is available for the depletants. The chemical potential of PHS depletants in the reservoir is simply given by the chemical potential of an ideal solution:

\begin{equation}
\mu_{\mathrm{d}}^{\mathrm{R}} = \mathrm{const} + k_{\mathrm{B}}T \, \mathrm{ln} \, n_{\mathrm{d}}^{\mathrm{R}}
\label{chemPotentialReservoir}
\quad\text{,}
\end{equation}

\noindent with $n_{\mathrm{d}}^{\mathrm{R}}$ the number density of depletants in the reservoir. By equating both expressions for the chemical potential of depletants, assuming equilibrium, an expression for $N_{\mathrm{d}}$ is found:

\begin{equation}
N_{\mathrm{d}} = n_{\mathrm{d}}^{\mathrm{R}} \, \langle V_{\mathrm{free}} \rangle
\quad\text{.}
\label{exprforNd}
\end{equation}

\noindent Substituting Eq. \ref{exprforNd} into Eq. \ref{semigrandpot2} and applying the Gibbs--Duhem relation:

\begin{equation}
\label{GDeq}
n_{\mathrm{d}}^{\mathrm{R}} \, \mathrm{d} \mu_{\mathrm{d}} = \mathrm{d} \Pi^{\mathrm{R}}
\quad\text{,}
\end{equation}

\noindent yields the following expression for the semi-grand potential of the system: 

\begin{equation}
\Omega(N_{\mathrm{c}}, V, T, \mu_{\mathrm{d}}) = F_{0}(N_{\mathrm{c}}, V, T) - \int\displaylimits_{0}^{\Pi^{\mathrm{R}}} \langle V_{\mathrm{free}} \rangle \, \mathrm{d} \Pi'^{\mathrm{R}} \quad\text{,}
\label{semigrandpot3}
\end{equation}

\noindent where $\Pi^{\mathrm{R}}$ is the osmotic pressure of depletants in the reservoir. Finally it is assumed that the PHS depletants do not influence the configuration of the colloids in the system. This implies that the free volume available for the depletants is equal to the free volume for depletants in the pure hard-sphere system; $\langle V_{\mathrm{free}} \rangle = \langle V_{\mathrm{free}} \rangle_{0}$. This leads to the following approximate result for the semi-grand potential of a mixture of hard spheres and penetrable hard spheres:

\begin{equation}
\Omega(N_{\mathrm{c}}, V, T, \mu_{\mathrm{d}}) = F_{0}(N_{\mathrm{c}}, V, T) - \langle V_{\mathrm{free}} \rangle_{0} \,  \Pi^{\mathrm{R}}
\quad\text{.}
\label{semigrandpot4}
\end{equation}

\noindent To compute phase equilibria it is useful to rewrite this equation in terms of dimensionless quantities as: 

\begin{equation}
\widetilde{\Omega} = \widetilde{F}_{0} - \alpha \, \widetilde{\Pi}^{\mathrm{R}}
\quad\text{,}
\label{semigrandpot5}
\end{equation}

\noindent where the following definitions are used:

\begin{multline}
\widetilde{\Omega} = \frac{\Omega \, v_{\mathrm{c}}}{V \, k_{\mathrm{B}} T} \quad\text{,}\quad\widetilde{F} = \frac{F \, v_{\mathrm{c}}}{V \, k_{\mathrm{B}} T} \quad\text{,} \\
\alpha = \frac{\langle V_{\mathrm{free}} \rangle_{0}}{V} \quad\text{,}\quad\widetilde{\Pi} = \frac{\Pi \, v_{\mathrm{c}}}{k_{\mathrm{B}} T} \quad\text{.}
\label{normalizedQuantities}
\end{multline}

\noindent Here $v_{\mathrm{c}}$ denotes the volume of a colloidal sphere. Furthermore, the size ratio $q$ is defined as the ratio of the radii of the depletants and colloids, $q = R_{\mathrm{d}} / R_{\mathrm{c}}$. Since the depletants do not have interactions with each other (i.e. ideal depletants), the osmotic pressure $\widetilde{\Pi}^{\mathrm{R}}$ is given by (the dimensionless form of) the van 't Hoff equation:

\begin{equation}
\widetilde{\Pi}^{\mathrm{R}}_{\mathrm{ideal}} = q^{-3} \, \phi_{\mathrm{d}}^{\mathrm{R}}
\quad\text{.}
\end{equation}

\noindent Here, $\phi_{\mathrm{d}}^{\mathrm{R}}$ is the volume fraction of depletants in the reservoir given by $n_{\mathrm{d}}^{\mathrm{R}} v_{\mathrm{d}}$. Furthermore, the Helmholtz free energy $\widetilde{F}_{0}$ of the pure hard-sphere fluid phase is given by:

\begin{equation}
\widetilde{F}_{0 \mathrm{,fluid}} = \phi_{\mathrm{c}} \, \left[  \mathrm{ln} \left( \phi_{\mathrm{c}} \, \Lambda^{3} / v_{\mathrm{c}} \right) - 1 \right] 
+ \frac{4\phi_{\mathrm{c}}^{2} - 3\phi_{\mathrm{c}}^{3}}{(1 - \phi_{\mathrm{c}})^{2}}
\quad\text{,}
\label{F0fluid}
\end{equation}

\noindent where the first term on the right-hand side is the ideal contribution and the second term originates from the Carnahan--Starling equation of state \cite{Carnahan1969}. For the hard-sphere solid, a face-centered-cubic (FCC) crystal, the result from Lennard-Jones and Devonshire cell theory \cite{Lennard1937} is used:

\begin{multline}
\widetilde{F}_{0 \mathrm{,solid}} = \phi_{\mathrm{c}} \, \mathrm{ln} \left( \Lambda^{3} / v_{\mathrm{c}} \right) +
\phi_{\mathrm{c}} \, \mathrm{ln} \left( \frac{27}{8 \phi_{\mathrm{cp}}^{3}} \right)\\
+ 3 \phi_{\mathrm{c}} \, \mathrm{ln} \left( \frac{\phi_{\mathrm{c}}}{1 - \phi_{\mathrm{c}}/\phi_{\mathrm{cp}}} \right)
\quad\text{.}
\label{F0solid}
\end{multline}

\noindent In Eqs. \ref{F0fluid} and \ref{F0solid}, $\Lambda$ is the De Broglie wavelength, $\phi_{\mathrm{c}}$ is the volume fraction of colloidal spheres, and $\phi_{\mathrm{cp}}$ denotes the volume fraction of a close-packed FCC crystal ($\phi_{\mathrm{cp}} = \pi / 3 \sqrt{2} \approx 0.74$). The final ingredient for the semi-grand potential $\widetilde{\Omega}$ in Eq. \ref{semigrandpot5} is the free volume fraction $\alpha$ which can be obtained from the reversible work $W$ required to insert a depletant into the system\cite{Lekkerkerker1992}:

\begin{equation}
\alpha = \exp\left[-\frac{W}{k_\text{B}T}\right]
\quad\text{.}
\label{alpha}
\end{equation}

\noindent The work of insertion $W$ is determined with scaled particle theory (SPT) \cite{Reiss1959}, resulting in:

\begin{multline}
\frac{W}{k_{\mathrm{B}} T} = -\mathrm{ln}(1-\phi_{\mathrm{c}}) + \frac{3 q \phi_{\mathrm{c}}}{1-\phi_{\mathrm{c}}} \\
+  \frac{1}{2} \left( \frac{6q^{2}\phi_{\mathrm{c}}}{1-\phi_{\mathrm{c}}} + \frac{9q^{2}\phi_{\mathrm{c}}^2}{(1-\phi_{\mathrm{c}})^2} \right) + q^{3} \, \widetilde{\Pi} \quad\text{.}
\label{Wresult}
\end{multline}

\noindent For the osmotic pressure $\widetilde{\Pi}$ in the last term on the right-hand side of Eq. \ref{Wresult}, the Carnahan--Starling equation \cite{Carnahan1969} is used for the fluid phase:

\begin{equation}
\widetilde{\Pi}_{\mathrm{fluid}} = \frac{\phi_{\mathrm{c}} + \phi_{\mathrm{c}}^{2} + \phi_{\mathrm{c}}^{3} - \phi_{\mathrm{c}}^{4}}{(1 - \phi_{\mathrm{c}})^{3}}
\quad\text{.}
\label{osmoticPressureFluid}
\end{equation}

\noindent For the solid phase, the osmotic pressure as derived from cell theory for an FCC crystal \cite{Lennard1937} is used:

\begin{equation}
\widetilde{\Pi}_{\mathrm{solid}} = \frac{3 \, \phi_{\mathrm{c}}}{1 - \phi_{\mathrm{c}}/\phi_{\mathrm{cp}}}
\quad\text{.}
\label{osmoticPressureSolid}
\end{equation}

\noindent It is noted here that the Percus--Yevick osmotic pressure (Eq. \ref{respress}) was used in original FVT \cite{Lekkerkerker1992}, as this osmotic pressure is internally consistent with SPT, both for the fluid phase and the solid phase.

The free volume fraction in the solid phase can also be determined using geometrical arguments \cite{Garcia2018a}, which gives a more accurate result for the solid phase at high volume fractions:

\begin{equation}
    \alpha_{\mathrm{solid}} = 
        \begin{cases}
            1-\phi_\text{c} \, \widetilde{v}_{\mathrm{exc}}^{0} & \text{for $\phi_\text{c} < \phi_\text{c}^{*}$ }\\
            1-\phi_\text{c} \, \widetilde{v}_{\mathrm{exc}}^{*} & \text{for $\phi_\text{c}^{*} \leq \phi_\text{c} < 2^{3/2} \, \phi_\text{c}^{*}$}\\
            0 & \text{otherwise}\quad\text{.}
        \end{cases}
    \label{geometricAlphaSolid}
\end{equation}

\noindent Here it is assumed that the centers of the spherical colloids are perfectly located on the FCC lattice points. In this equation, $\phi_\text{c}^{*} = \phi_\text{c}^{\mathrm{cp}} \, / \,\widetilde{v}_{\mathrm{exc}}^{0}$ denotes the volume fraction of large spheres above which the depletion zones overlap. Furthermore, the normalized excluded volumes are given by:

\begin{equation}
\widetilde{v}_{\mathrm{exc}}^{0} = (1 + q)^{3}
\quad\text{,} 
\end{equation}
\begin{multline}
\widetilde{v}_{\mathrm{exc}}^{*} = \widetilde{v}_{\mathrm{exc}}^{0} - 6 \left[ 1 + q - \left( \frac{\phi_\text{c}^{\mathrm{cp}}}{\phi_\text{c}} \right)^{\frac{1}{3}} \right]^{2} \\
\left[ 1 + q + \frac{1}{2} \left( \frac{\phi_\text{c}^{\mathrm{cp}}}{\phi_\text{c}} \right)^{\frac{1}{3}} \right]
\quad\text{.}
\end{multline}

\noindent Note that Eq. \ref{geometricAlphaSolid} only holds if there is no multiple overlap of depletion zones, i.e. for size ratios smaller than $q = \frac{2}{3}\sqrt{3}-1\approx0.15$. 

\subsubsection{First $\Omega$ correction for HS depletants}
\noindent The semi-grand potential $\Omega$ can also be used to describe a mixture of large and small spheres in equilibrium with a reservoir of small spheres. For this purpose, Lekkerkerker and Stroobants\cite{Lekkerkerker1993} proposed to use a different expression in Eq. \ref{semigrandpot5} for the osmotic pressure in the reservoir to account for the excluded volume interactions between the depletants. Instead of assuming ideal behaviour, the "compressibility" result of the Percus-Yevick (PY) clossure was used, which describes the osmotic pressure of monodisperse hard-sphere dispersions with reasonable accuracy:

\begin{equation}
\widetilde{\Pi}^{\mathrm{R}}_{\mathrm{PY}} = q^{-3} \, \frac{\phi_{\mathrm{d}}^{\mathrm{R}} + (\phi_{\mathrm{d}}^{\mathrm{R}})^{2} + (\phi_{\mathrm{d}}^{\mathrm{R}})^{3}}{(1 - \phi_{\mathrm{d}}^{\mathrm{R}})^{3}}
\quad\text{.}
\label{respress}
\end{equation}

\begin{figure}[tb!]
	\centering
		\includegraphics[width=.48\textwidth]{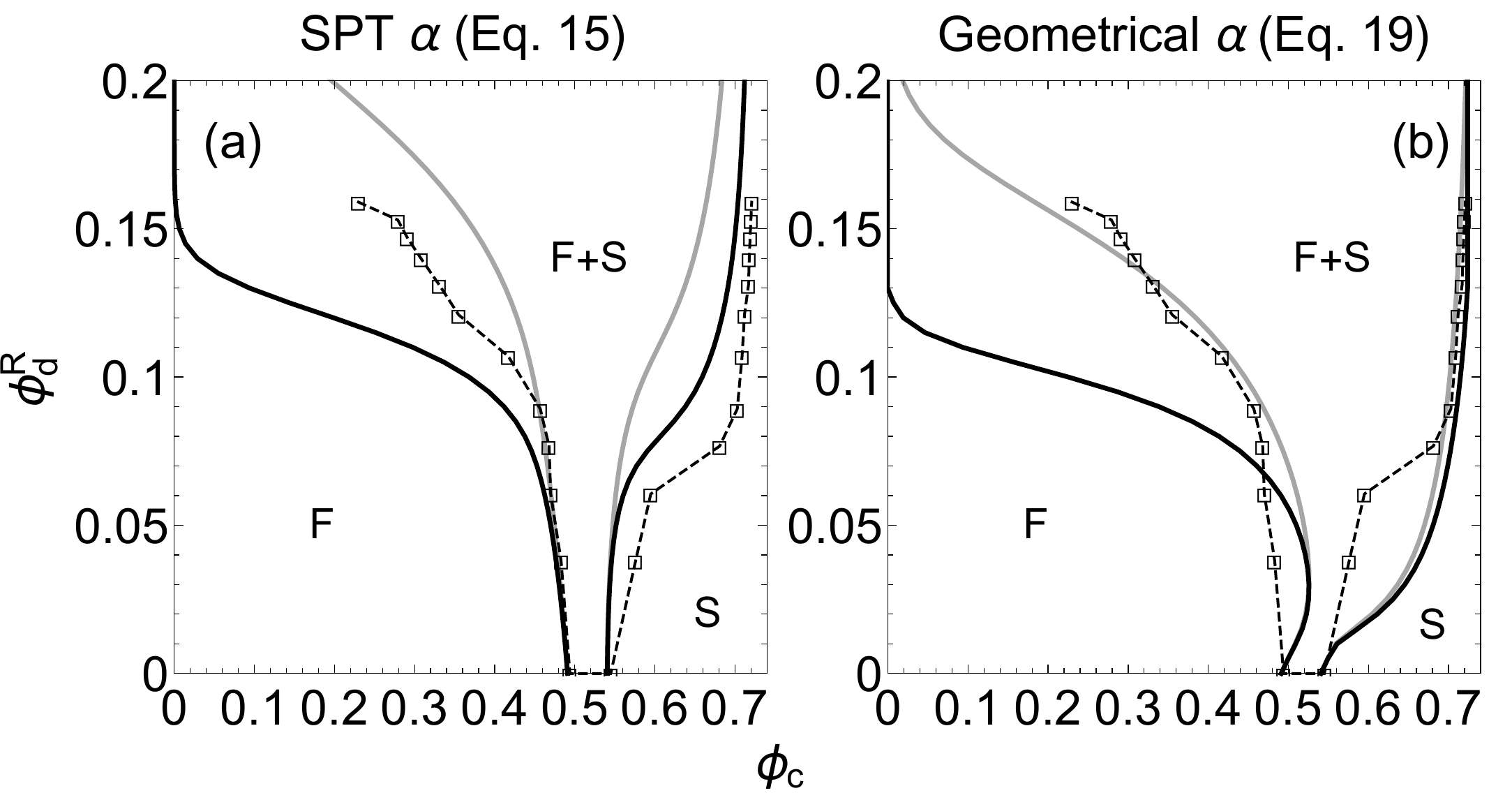}
	\caption{Comparison between the phase diagrams of a binary hard-sphere mixture with size ratio $q = 0.1$ determined with FVT using the PHS approximation \cite{Lekkerkerker1992} (gray curves), FVT for hard-sphere depletants \cite{Poon1994,Lekkerkerker2002} (black curves) and fluid-solid coexistence lines from direct coexistence simulations \cite{Dijkstra1999} (open square symbols, the dashed lines are a guide for the eyes). In contrast to the theoretical results, simulations also revealed a small region of solid-solid coexistence at high concentrations of large spheres, but these results are omitted here. The free volume fraction obtained from SPT (Eq. \ref{alpha}) is used for the FVT calculations in (a) and the FVT results for the geometrical expression for the free volume fraction (Eq. \ref{geometricAlphaSolid}) are shown in (b). The fluid phase is denoted with F, the solid phase with S and the coexistence region with F+S. }   
	\label{fig2}
\end{figure}

Fig. \ref{fig2} shows the phase diagram for a hard-sphere mixture with size ratio $q = 0.1$ determined using original FVT for hard-sphere depletants \cite{Lekkerkerker1993,Poon1994,Lekkerkerker2002} compared with direct coexistence simulation results of Dijkstra \textit{et al.} \cite{Dijkstra1999} The results are shown both for the SPT expression (Eq. \ref{alpha}) and the geometrical expression for the free volume fraction in the solid phase (Eq. \ref{geometricAlphaSolid}). Also shown for comparison are the binodals obtained with FVT using the PHS approximation (gray curves). The SPT expression was originally used in FVT for hard-sphere depletants and the resulting binodals are in slightly better agreement with simulation results than the binodals for PHS depletants. However, still a significant discrepancy remains and also when the geometrical expression for $\alpha_\text{solid}$ is used this discrepancy is present. The reason for this mismatch is that the excluded volume of the depletants is still not explicitly taken into account in the FVT of Lekkerkerker and Stroobants \cite{Lekkerkerker1993}. The excluded volume of the depletants is not accounted for in the free volume fraction, for both the reservoir and the system, and therefore the chemical potential of the depletants is not accurately taken into account. The chemical potential for depletants in the system given by Eq. \ref{chemPotentialSystem} still holds for hard-sphere depletants, but the volume excluded by the depletants has to be accounted for properly in the free volume $\langle V_\text{free}\rangle$. The chemical potential for depletants in the reservoir given by Eq. \ref{chemPotentialReservoir} is no longer valid since hard spheres at finite concentrations do not behave ideally. Moreover, the approximation used in Eq. \ref{semigrandpot4} is no longer valid because the free volume available to the depletants is no longer independent of depletant concentration. In the next subsection we derive an adjusted expression for the semi-grand potential of binary hard-sphere mixtures that accounts for the excluded volume of the depletants more accurately. 

It is noted that the binodals of FVT for PHS depletants in Fig. \ref{fig2}b are in remarkable agreement with the simulation results of the binary hard-sphere mixture, except for the low depletant concentration region. This similarity in phase behaviour for PHS depletants and hard-sphere depletants for large size discrepancies ($q \lesssim 0.2$) was also found by Velasco \textit{et al.}\cite{Velasco1999}. Using a perturbation theory they determined the phase behaviour of a colloidal mixture using a variety of different model pair potentials. It was found that the exact shape of the depletion potential barely affects the phase behaviour of the system as various hard-sphere pair potentials yield essentially the same phase diagram as the Asakura--Oosawa pair potential. Moreover, a mismatch between simulation and theoretical results in the region of low depletant concentrations and high colloid concentrations was found by Velasco \textit{et al.} for both the Asakura--Oosawa pair potential and the hard-sphere pair potentials, similar to the mismatch in FVT that can be seen in Fig. \ref{fig2}b. A downside of the perturbation theory is that it relies on a pair potential to account for depletion, which makes it difficult to apply to colloidal mixtures with large size discrepancies or containing anisotropic depletants, whereas FVT does not use a pair potential but is solely based on the free volume available to the depletants.

\subsubsection{Adjusted $\Omega$ for HS depletants}\label{sssec213}
\noindent Next, the excluded volume of hard-sphere depletants is accounted for in both the reservoir and the system. We start from the definition of the semi-grand potential given by Eq. \ref{semigrandpot2}. An equation for the number of small hard-sphere depletants in the system is again obtained by equating the chemical potentials of the small spheres in the system and in the reservoir. Non-ideal behaviour can be accounted for in the chemical potential of the small spheres, both in the system and in the reservoir, by the work of small sphere-insertion:

\begin{equation}
\label{HSmuS}
\widetilde{\mu}_\text{d} = \mathrm{const} + \mathrm{ln}(\phi_\text{d}) + \frac{W^\text{S}}{k_{\mathrm{B}}T}
\quad\text{,}
\end{equation}
\begin{equation}
\label{HSmuR}
\widetilde{\mu}_\text{d}^\text{R} = \mathrm{const} + \mathrm{ln}(\phi_\text{d}^\text{R}) + \frac{W^\text{R}}{k_{\mathrm{B}}T}
\quad\text{,}
\end{equation}

\noindent where $W^\text{S}$ is now the work of inserting a hard-sphere depletant in the system consisting of a binary sphere mixture, and $W^\text{R}$ is the work of inserting a hard-sphere depletant in the reservoir, which is a dispersion containing only hard-sphere depletants. Combining Eq. \ref{alpha} with Eqs. \ref{HSmuS} and \ref{HSmuR}, the following expressions are found for the volume fraction of depletants in the system $\phi_\text{d}$ for the fluid phase and solid phase, respectively:

\begin{equation}
\phi_\text{d} = \phi_\text{d}^{\mathrm{R}} \, \frac{\alpha_{\mathrm{fluid}}^\text{S}(q, \phi_\text{c}, \phi_\text{d})}{\alpha^{\mathrm{R}}(\phi_\text{d}^{\mathrm{R}})}
\label{eta2mixturefluid}
\quad\text{,}
\end{equation}
\begin{equation}
\phi_\text{d} = \phi_\text{d}^{\mathrm{R}} \, \frac{\alpha_{\mathrm{solid}}^\text{S}(q, \phi_\text{c}, \phi_\text{d})}{\alpha^{\mathrm{R}}(\phi_\text{d}^{\mathrm{R}})}
\label{eta2mixturesolid}
\quad\text{.}
\end{equation}

\noindent In Eqs. \ref{eta2mixturefluid} and \ref{eta2mixturesolid}, it is stressed that the free volume fraction in the reservoir is no longer unity and $\alpha_{\mathrm{fluid}}^\text{S}$ and $\alpha_{\mathrm{solid}}^\text{S}$ do not only depend on the volume fraction of large spheres $\phi_\text{c}$, but also on the volume fraction of the depletants. The volume fraction of depletants in the system $\phi_\text{d}$, in coexistence with the reservoir with a certain depletant volume fraction $\phi_\text{d}^\text{R}$, can be found numerically by solving Eqs. \ref{eta2mixturefluid} and \ref{eta2mixturesolid}. Substituting Eqs. \ref{eta2mixturefluid} and \ref{eta2mixturesolid} into the definition of the semi-grand potential given by Eq. \ref{semigrandpot2} and applying the Gibbs--Duhem relation (Eq. \ref{GDeq}) finally yields expressions for the semi-grand potential for the fluid and solid phases of a binary hard-sphere mixture:

\begin{equation}
\widetilde{\Omega}_{\mathrm{fluid}} = \widetilde{F}_{0 \mathrm{,fluid}} - \int\displaylimits_{0}^{\phi_\text{d}^{\mathrm{R}}} \frac{\alpha_{\mathrm{fluid}}^\text{S}(q, \phi_\text{c}, \phi_\text{d})}{\alpha^{\mathrm{R}}(\phi_\text{d}^{\mathrm{R}})} \, \left( \frac{\partial \widetilde{\Pi}^{\mathrm{R}}}{\partial \phi_\text{d}^{\mathrm{R}'}} \right) \, \mathrm{d} \phi_\text{d}^{\mathrm{R}'}
\label{endOmegaFluid}
\quad\text{,}
\end{equation}
\begin{equation}
\widetilde{\Omega}_{\mathrm{solid}} = \widetilde{F}_{0 \mathrm{,solid}} - \int\displaylimits_{0}^{\phi_\text{d}^{\mathrm{R}}} \frac{\alpha_{\mathrm{solid}}^\text{S}(q, \phi_\text{c}, \phi_\text{d})}{\alpha^{\mathrm{R}}(\phi_\text{d}^{\mathrm{R}})} \, \left( \frac{\partial \widetilde{\Pi}^{\mathrm{R}}}{\partial \phi_\text{d}^{\mathrm{R}'}} \right) \, \mathrm{d} \phi_\text{d}^{\mathrm{R}'}
\label{endOmegaSolid}
\quad\text{,}
\end{equation}

\noindent where the dimensionless quantities from Eq. \ref{normalizedQuantities} are applied and the integration variable d$\mu'_{\mathrm{d}}$ in Eq. \ref{semigrandpot2} is changed to the volume fraction of depletants in the reservoir d$\phi_\text{d}^{\mathrm{R}'}$. The free volume fraction for depletants in the reservoir $\alpha^\text{R}$ can be calculated by using Eqs. \ref{alpha}-\ref{osmoticPressureFluid} and using $q = 1$. It is noted that Eqs. \ref{endOmegaFluid} and \ref{endOmegaSolid} recover the original semi-grand potential given by Eq. \ref{semigrandpot5} when the PHS approximation is applied. In the next section we discuss how to obtain expressions for the free volume fraction of hard-sphere depletants in the binary system for both the fluid and solid phase. Even though the excluded volume of the small spheres will be accounted for in the free volume fractions, it is still assumed that the presence of the depletants does not alter the configurations of the large spheres in our approach outlined below.

\subsection{Free volume fraction for HS depletants} \label{ssec22}

\subsubsection{Fluid phase}
\noindent The free volume fraction in the fluid phase of a binary hard-sphere mixture can again be determined using the work for depletant insertion in a binary mixture given by SPT\cite{Lebowitz1965}, resulting in:

\begin{multline}
\alpha^{\mathrm{S}}_{\mathrm{fluid}} = \exp-\bigg[-\mathrm{ln}(1 - \phi_\text{c} - \phi_\text{d}) + \frac{3 q \phi_\text{c} + 3 \phi_\text{d}}{1 - \phi_\text{c} - \phi_\text{d}} + \frac{3 q \phi_\text{c} + 3 \phi_\text{d}}{1 - \phi_\text{c} - \phi_\text{d}} \\ + \frac{1}{2} \left(\frac{6 q^{2} \phi_\text{c} + 6 \phi_\text{d}}{1 - \phi_\text{c} - \phi_\text{d}} + \left( \frac{3 q \phi_\text{c} + 3 \phi_\text{d}}{1 - \phi_\text{c} - \phi_\text{d}} \right)^{2} \right) +q^{3}\widetilde{\Pi}_{\mathrm{BM}}\bigg]
\quad\text{,}
\label{insertionworkHSdepletant}
\end{multline}

\noindent where $\widetilde{\Pi}_{\mathrm{BM}}$ is the osmotic pressure of the binary mixture of hard spheres. An expression for $\widetilde{\Pi}_{\mathrm{BM}}$ is given by the Boublik--Mansoori--Carnahan--Starling--Leland (BMCSL) equation of state for binary hard-sphere mixtures \cite{Boublik1970,Mansoori1971}:

\begin{equation}
\begin{aligned}
&\widetilde{\Pi}_{\mathrm{BM}} = \frac{\phi_\text{c}+ q^{-3}\phi_\text{d}}{1 - \phi_\text{c} - \phi_\text{d}} \\ &+ 3 \frac{\phi_\text{c}^{2} + q^{-1} \phi_\text{c}\phi_\text{d} + q^{-2} \phi_\text{c}\phi_\text{d} + q^{-3} \phi_\text{d}^{2}}{(1 - \phi_\text{c} - \phi_\text{d})^{2}} \\ &+ \left(\frac{\phi_\text{c}^{3} + 3 q^{-1} \phi_\text{c}^{2}\phi_\text{d} + 3 q^{-2} \phi_\text{c}\phi_\text{d}^{2} + q^{-3} \phi_\text{d}^{3}}{(1 - \phi_\text{c} - \phi_\text{d})^{3}}\right) \\ 
&\times\left(3 - \phi_\text{c} - \phi_\text{d}\right)
\quad\text{.}
\end{aligned}
\end{equation}

\subsubsection{Solid phase ($q \lesssim 0.2$)}
\noindent The same approach cannot be followed for the solid phase since the osmotic pressure of a hard-sphere solid containing smaller hard spheres is not known. Moreover, as mentioned in Sec. \ref{sssec:PHS}, the scaled particle theory approach does not accurately describe the free volume available in the solid phase at high concentrations\cite{Garcia2018a}. The free volume fraction in the solid phase is approximated here by considering an FCC crystal of the larger spheres and assuming that the small spheres behave as a fluid in the free space left by the large spheres, which is valid for highly asymmetric binary sphere mixtures\cite{Filion2011,Dijkstra1999} with $q \lesssim 0.2$. With this assumption, the free volume fraction $\alpha_\text{solid}^\text{S}$ can be approximated by a product of the free volume fraction of the hard-sphere solid and the free volume fraction in the small sphere fluid that surrounds the larger spheres: 

\begin{equation}
\alpha^{\mathrm{S,approx}}_{\mathrm{solid}}(q,\phi_\text{c},\phi_\text{d}) = \alpha_{\mathrm{solid}}(q, \phi_\text{c}) \, \alpha_{\mathrm{fluid}}(q=1, \phi_\text{d}^{\dagger}) 
\quad\text{,}
\label{alphamixs}
\end{equation}

\noindent where $\phi_\text{d}^{\dagger} = \phi_\text{d} / (1-\phi_\text{c})$ is the effective volume fraction of the small spheres in the space that is not occupied by large spheres and $\alpha_{\mathrm{solid}}(q, \phi_\text{c})$ is given by the geometrical free volume fraction in Eq. \ref{geometricAlphaSolid}. This expression is only a rough approximation, as it does not accurately account for the overlap between the depletion zones of large and small spheres. To take this overlap into account, we make the same approximation for the fluid phase and use the ratio of this approximation and $\alpha^\text{S}_\text{fluid}$ from SPT given by Eq. \ref{insertionworkHSdepletant} as a correction factor that takes the overlap between the depletion zones of small and large spheres into account:

\begin{equation}
\alpha^{\mathrm{S,approx}}_{\mathrm{fluid}}(q,\phi_\text{c},\phi_\text{d}) = \alpha_{\mathrm{fluid}}(q, \phi_\text{c}) \, \alpha_{\mathrm{fluid}}(q=1, \phi_\text{d}^{\dagger}) 
\quad\text{,}
\label{alphamixf}
\end{equation}
\begin{multline}
\begin{aligned}
\alpha^{\mathrm{S}}_{\mathrm{solid}}(q,\phi_\text{c},\phi_\text{d}) &= \alpha^{\mathrm{S,approx}}_{\mathrm{solid}}(q,\phi_\text{c},\phi_\text{d})\frac{\alpha^\text{S}_\text{fluid}(q,\phi_\text{c},\phi_\text{d})}{\alpha^{\mathrm{S,approx}}_{\mathrm{fluid}}(q,\phi_\text{c},\phi_\text{d})}\\
&= \alpha_{\mathrm{solid}}(q, \phi_\text{c})\frac{\alpha^{\mathrm{S}}_{\mathrm{fluid}}(q,\phi_\text{c},\phi_\text{d})}{\alpha_{\mathrm{fluid}}(q, \phi_\text{c})}
\quad\text{.}
\label{alphamixs2}
\end{aligned}
\end{multline}

\noindent The result in equation \ref{alphamixs2} implies that the ratio between the free volume for a depletant in the binary system (denoted as $\alpha^\text{S}$) and in a system with only large particles is independent of the phase of the large particles.
In Sec. \ref{alpha results} it is shown that Eq. \ref{alphamixs2} accurately matches simulation data for dense colloidal hard-sphere mixtures with size ratios $q = 0.1$ and $q = 0.05$. 

\subsubsection{Solid phase ($ q > 0.2$)} \label{sssec:223}
\noindent The analysis in the previous section only holds when the small spheres behave as a fluid in the FCC crystal of the large particles, but for $q>0.2$ this is no longer the case and at these size ratios either an interstitial solid solution or a cocrystal is formed \cite{Filion2009,Filion2011,Filion2011a}. Due to the large variety of crystal structures that can be formed for hard-sphere mixtures with larger size ratios\cite{Sanders1978,Leunissen2005,Filion2009}, a general FVT cannot be derived because each different cocrystal requires an accurate thermodynamic description and free volume fraction. However, for specific size ratios FVT might still be applicable. The Monte Carlo approach of Filion and Dijkstra\cite{Filion2009} can be used to predict the crystal structures of binary mixtures with a specific size ratio. This may be used to obtain the required input for an accurate FVT approach for specific mixtures. Here we focus on the case of a mixture of hard spheres with size ratio $q=0.4$. At this size ratio the binary mixture forms an interstitial solid solution at high densities \cite{Filion2009,Filion2011}. In this relatively simple binary system, the large spheres organize in an FCC structure where the small spheres do not fit in the tetrahedral holes and there is space for one small sphere in the octahedral holes of the FCC crystal formed by the large particles.

A new expression is needed for the free volume available in the FCC crystal formed by the hard spheres $\alpha_\text{solid}$ since Eq. \ref{geometricAlphaSolid} can no longer be applied for $q = 0.4$ due to multiple overlap of depletion zones. Finding an analytical equation for $\alpha_\text{solid}$ for $q > 0.15$ is more difficult since already accounting for overlap of three depletion zones is mathematically laborious \cite{Chkhartishvili2001}. However, it is possible to find an equation for the free volume available in an FCC crystal, or any other given (binary) crystal structure, accounting for multiple overlaps with a numerical approach. One way to do this is using Wolfram Mathematica's built-in Region functions\cite{numalpha}. The free volume fraction of the one component FCC crystal obtained for $q = 0.4$ following this approach\cite{numalpha} is given by:

\begin{multline}
\label{alfa04}
\alpha_{\mathrm{solid}}^\text{fit}\left(q = 0.4,\phi_\text{c}\right) = 1.33-9.38\phi_\text{c}+28.5\phi_\text{c}^2- \\48.0\phi_\text{c}^3+
47.1\phi_\text{c}^4-25.5\phi_\text{c}^5+5.97\phi_\text{c}^6
\quad\text{.}
\end{multline}

This result is used to determine the free volume fraction of the interstitial solid solution. Given that the number of octahedral holes in the FCC crystal of the large particles is the same as the number of large particles, and assuming that the excluded volume of a small sphere present in an octahedral hole completely fills the hole, the total free volume fraction in the system can be described as:

\begin{equation}\label{alpha04mix}
\alpha^{\mathrm{S}}_{\mathrm{solid}}(q=0.4,\phi_\text{c},\phi_\text{d}) = \alpha_{\mathrm{solid}}^\text{fit}\left(q = 0.4,\phi_\text{c}\right)\left(1-q^{-3}\frac{\phi_\text{d}}{\phi_\text{c}}\right)
\quad\text{,}
\end{equation} 

\noindent where the term $q^{-3}(\phi_\text{d}/\phi_\text{c})$ can be interpreted as a filling fraction of small spheres in the octahedral holes.

\subsection{Phase coexistence calculations}
\noindent Two-phase coexistence densities of a system containing hard spheres and depletants are determined by applying the coexistence criteria of an equilibrium between a phase I and a phase II:

\begin{equation}
\widetilde{\mu}^{\mathrm{I}}_{\mathrm{c}} = \widetilde{\mu}^{\mathrm{II}}_{\mathrm{c}}
\quad\text{,}\quad
\widetilde{\Pi}^{\mathrm{I}} = \widetilde{\Pi}^{\mathrm{II}}
\quad\text{.}
\label{eqcrit1}
\end{equation}

\noindent The chemical potential of the large spheres and the osmotic pressure are calculated with the following thermodynamic relations:

\begin{equation}
\widetilde{\mu}_{\mathrm{c}} = \left( \frac{\partial \widetilde{\Omega}}{\partial \phi_{\mathrm{c}}} \right)_{\widetilde{\Pi}^{\mathrm{R}},V,T}
\quad\text{,}
\label{derivativeOmega}
\end{equation}
\begin{equation}
\widetilde{\Pi} = \phi_{\mathrm{c}} \, \widetilde{\mu}_{\mathrm{c}} - \widetilde{\Omega}
\quad\text{.}
\label{osmoticPresRelation}
\end{equation}

\noindent Note that the equalities given by Eq. \ref{eqcrit1} correspond to a common tangent construction applied on the semi-grand potential $\widetilde{\Omega}$ as a function of $\phi_\text{c}$ with slope $\widetilde{\mu}_{\mathrm{c}}$ and intercept $-\widetilde{\Pi}$. Numerical expressions for the semi-grand potential in the fluid and solid phase of the binary hard-sphere mixture were obtained according to the following procedure. First, the depletant concentration and free volume fraction in the system are determined by solving Eqs. \ref{eta2mixturefluid} and \ref{eta2mixturesolid} for different values of $\phi_\text{d}^\text{R}$ and a given $\phi_\text{c}$. This data is then fitted using interpolation and used as input for Eqs. \ref{endOmegaFluid} and \ref{endOmegaSolid} to determine $\widetilde{\Omega}$ for a given $\phi_\text{c}$. This is repeated for different values of $\phi_\text{c}$ and again interpolation is used to get an expression for $\widetilde{\Omega}$ as a function of $\phi_\text{c}$ for a given $\phi_\text{d}^\text{R}$ and $q$. Binodals were finally determined by solving Eqs. \ref{eqcrit1}-\ref{osmoticPresRelation} with Mathematica's built-in FindRoot function and repeating this procedure for a range of reservoir concentrations $\phi_{\mathrm{d}}^{\mathrm{R}}$. The phase diagrams were converted from the $(\phi_{\mathrm{c}},\phi_{\mathrm{d}}^{\mathrm{R}})$-plane into the $(\phi_{\mathrm{c}},\phi_{\mathrm{d}})$-plane by making use of Eqs. \ref{eta2mixturefluid} and \ref{eta2mixturesolid}.

\begin{figure}[tb!]
	\centering
		\includegraphics[width=.48\textwidth]{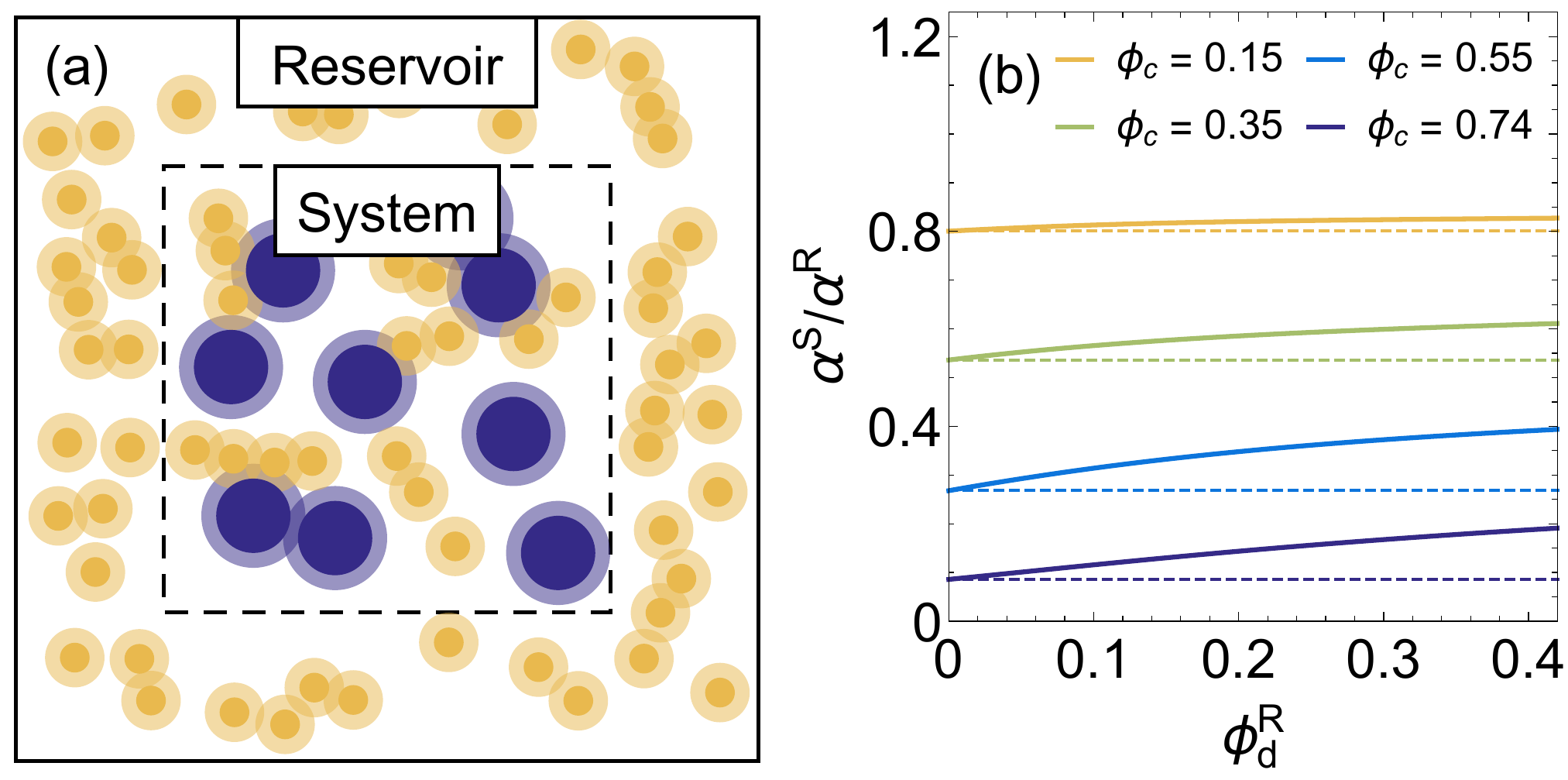}
	\caption{(a) Schematic representation of the FVT approach for hard-sphere depletants explained in Sec. \ref{sssec213}. The dashed black lines represent the semi-permeable membrane that the large colloidal spheres cannot pass through, but is permeable to the small spheres and solvent. The semi-transparent regions around the small and large spheres indicate the depletion zones that are inaccessible to the centers of the small spheres. The white area shows the free volume that is available to the depletants. The size ratio between the colloidal spheres is 0.4. (b) The ratio of the free volumes in the system and in the reservoir $\alpha^\text{S}/\alpha^\text{R}$ as a function of the depletant volume fraction in the reservoir $\phi_\text{d}^\text{R}$ for different volume fractions of the large spheres $\phi_\text{c}$ and a size ratio of $q = 0.1$. The solid curves are obtained from the theory presented in Sec. \ref{ssec22} with $\alpha^\text{S}$ given by Eq. \ref{insertionworkHSdepletant} for the fluid phase ($\phi_\text{c}=0.15$ and $\phi_\text{c}=0.35$) and Eq. \ref{alphamixs2} for the solid phase ($\phi_\text{c}=0.55$ and $\phi_\text{c}=0.74$). The dashed curves indicate the results for the PHS approximation, with Eq. \ref{geometricAlphaSolid} for $\alpha_\text{solid}$.}
	\label{fig3}
\end{figure}

\section{Results \& Discussion}
\noindent Here we present and discuss results of the theoretical method described in the previous section and verify the validity. First, we show how the adjusted description of the semi-grand potential that explicitly takes the excluded volume of the depletants into account deviates from the original semi-grand potential used in FVT for binary hard-sphere mixtures. Second, we test the validity of the expressions used for the free volume fraction $\alpha^\text{S}$ in the binary mixture by comparing the relation between the concentration of depletants in the reservoir and in the system with computer simulation results. Next, we show how multiple overlap influences the free volume fraction in the solid phase. Subsequently, we present phase diagrams for size ratios $q = 0.05, 0.1, 0.2$ and $0.4$ and make a comparison with phase diagrams obtained from simulations. Finally, possible extensions of our approach are discussed.

\subsection{Free volume fraction} \label{alpha results}
\noindent The main difference between the theory presented in Secs. \ref{sssec213} and \ref{ssec22} with respect to the original FVT for binary hard-sphere mixtures is that the excluded volume of the depletants is explicitly taken into account in the free volume descriptions. Due to this, the free volume available to the depletants in the system and in the reservoir is significantly lower. Fig. \ref{fig3}a shows a schematic picture of the FVT approach presented in this paper. Both the large and small colloidal particles are surrounded by a depletion zone that is inaccessible to the small particles and the white areas in both the system and the reservoir show the free volume available to the depletants. In original FVT the excluded volume of the depletants was not fully taken into account and as a result the free volume fraction is always unity in the reservoir and the free volume fraction in the system is independent of the depletant concentration. When the excluded volume interactions of the depletants are taken into account this is no longer the case, as presented in Sec. \ref{sssec213}, and both the free volume in the reservoir ($\alpha^\text{R}$) and the system ($\alpha^\text{S}$) depend on the depletant concentration. The effect of this excluded volume interaction is demonstrated in Fig. \ref{fig3}b where the ratio $\alpha^\text{S}/\alpha^\text{R}$, the relative fraction of the volume available in the system with respect to that available in the reservoir, is plotted as a function of the depletant volume fraction for different large particle concentrations $\phi_\text{c}$. This ratio is given by Eqs. \ref{eta2mixturefluid} and \ref{eta2mixturesolid} and is an important contribution to the semi-grand potential of the system given by Eqs. \ref{endOmegaFluid} and \ref{endOmegaSolid}. Fig. \ref{fig3}b shows that $\alpha^\text{S}/\alpha^\text{R}$ is no longer constant as originally assumed by Lekkerkerker and Stroobants\cite{Lekkerkerker1993} but increases as a function of the depletant concentration in the reservoir. The difference with respect to original FVT becomes more significant at higher concentrations of either the large or small hard spheres. 

\begin{figure}[tb!]
	\centering
		\includegraphics[width=.48\textwidth]{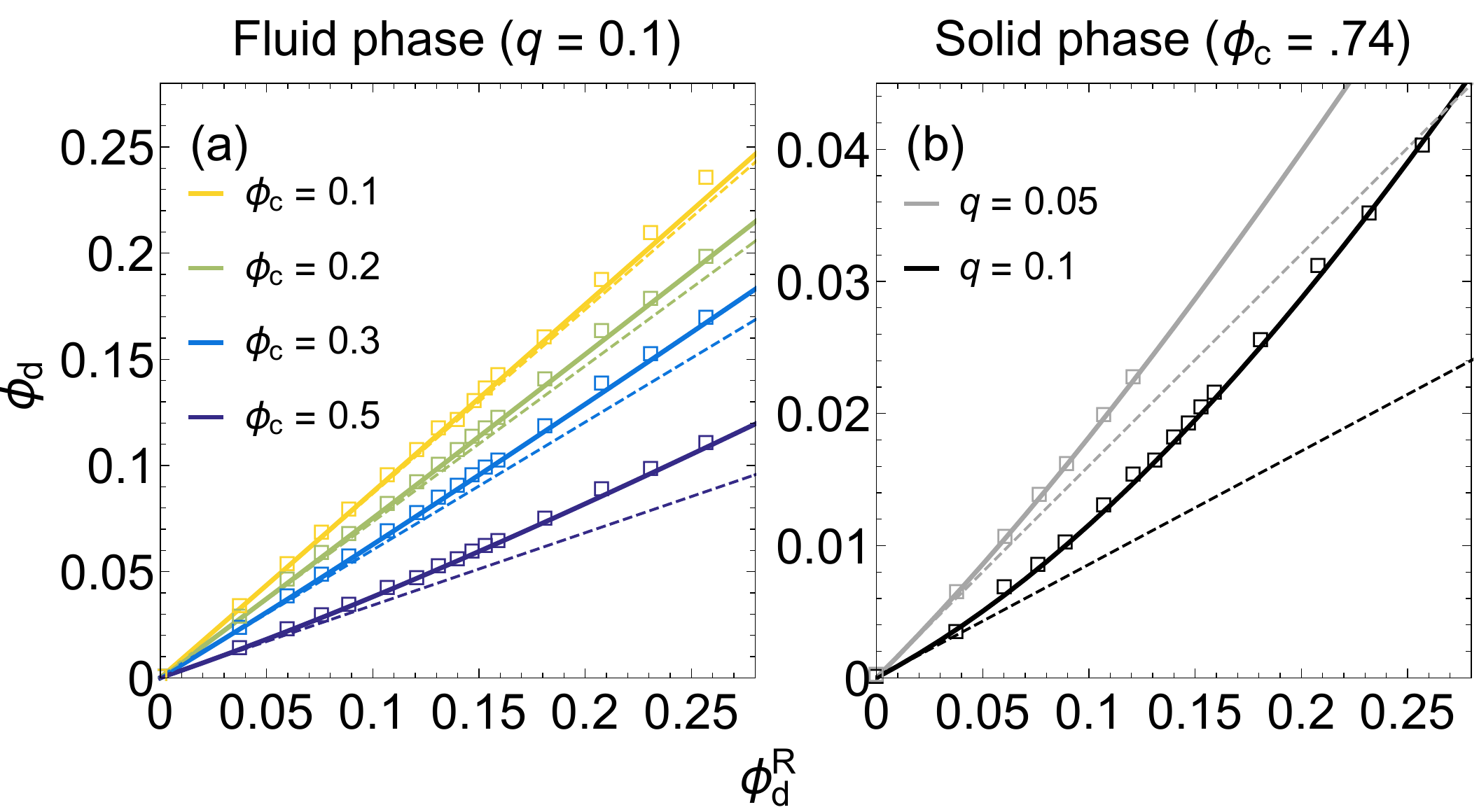}
	\caption{Relation between $\phi_\text{d}^\text{R}$ and $\phi_\text{d}$ as determined with Eqs. \ref{eta2mixturefluid} and \ref{eta2mixturesolid} (solid curves). The free volume fraction in the system $\alpha^\text{S}$ is given by Eq. \ref{insertionworkHSdepletant} for the fluid phase (a) and Eq. \ref{alphamixs2} for the solid phase (b). The values used for the volume fraction of large spheres $\phi_\text{c}$ and the size ratio $q$ are indicated in the figure. The dashed curves show the results for PHS depletants using Eq. \ref{geometricAlphaSolid} for $\alpha_\text{solid}$. The symbols denote Monte Carlo computer simulation data from Dijkstra \textit{et al.}\cite{Dijkstra1999}.}
	\label{fig4}
\end{figure}

Fig. \ref{fig4} shows the relation between the volume fraction of depletants in the system $\phi_\text{d}$ and in the reservoir $\phi_\text{d}^\text{R}$ for the fluid phase (a) and the solid phase (b). As mentioned above, this relation is given by the $\alpha^\text{S}/\alpha^\text{R}$ ratio. Also shown in Fig. \ref{fig4} is the relation from original FVT theory and computer simulation data by Dijkstra \textit{et al.} \cite{Dijkstra1999}. The results from the adjusted FVT follow the simulation data remarkably well, which confirms the validity of the equations obtained for the free volume fractions in the fluid phase and the solid phase given by Eqs. \ref{insertionworkHSdepletant} and \ref{alphamixs2}. The free volume approach followed here also compares strikingly well with an alternative relation between $\phi_\text{d}^\text{R}$ and $\phi_\text{c}$ derived by Roth \textit{et al.} \cite{Roth2001} using a density functional theory approach.   

\begin{figure}[tb!]
	\centering
		\includegraphics[width=.48\textwidth]{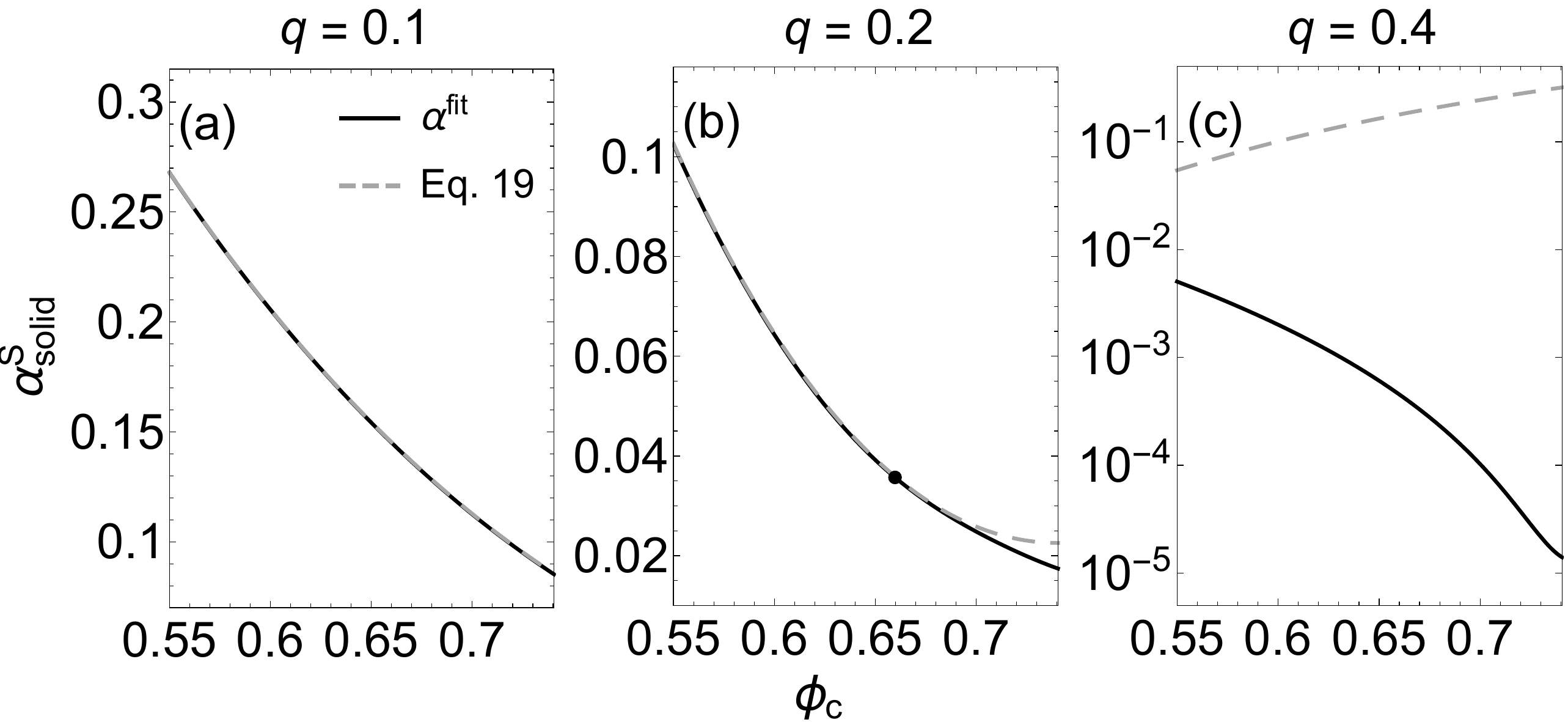}
	\caption{Free volume fraction based upon geometrical arguments for a PHS depletant in a colloidal FCC crystal with the colloidal particles fixed at their respective lattice positions. The solid black curves show the results for the numerical procedure explained in Sec. \ref{sssec:223} that takes multiple overlap of depletion zones into account. The dashed gray curves show the results of Eq. \ref{geometricAlphaSolid}, which is analytical but no longer holds when multiple overlap occurs\cite{Garcia2018a}. The size ratios considered are $q = 0.1$ where no multiple overlap occurs (a), $q = 0.2$ where multiple overlap occurs at high densities ($\phi_\text{c}\geq 0.66$) (b) and $q = 0.4$ where multiple overlap occurs for all concentrations where a solid phase is expected (c).}
	\label{fig5}
\end{figure}

As mentioned previously, the geometrical description of the free volume in a single component FCC crystal given by Eq. \ref{geometricAlphaSolid} is not valid anymore for $q > 0.15$ due to multiple overlap of depletion zones and a numerical method is used to obtain a description for $\alpha_\text{solid}$ as described in Sec. \ref{sssec:223}. Unfortunately, for these size ratios there is no simulation data on the free volume or on the equilibrium between $\phi_\text{d}$ and $\phi_\text{d}^\text{R}$ available in the literature for comparison. Fig. \ref{fig5} shows the results of the numerical free volume fraction in the solid phase ($\alpha^\text{fit}$) as a function of the volume fraction of the large particle solid $\phi_\text{c}$ for a depletant with size ratios $q = 0.1, 0.2$ and $0.4$. Also shown for comparison is Eq. \ref{geometricAlphaSolid}, which does not account for multiple overlap. The free volume $\alpha^\text{fit}$ matches with the analytical result of Eq. \ref{geometricAlphaSolid} for $q = 0.1$, as expected because no multiple overlap occurs for this size ratio. For $q = 0.2$ the numerical method also corresponds to Eq. \ref{geometricAlphaSolid} for low particle concentrations, however at a certain point multiple overlap of depletion zones occurs and the free volume fraction starts to deviate from Eq. \ref{geometricAlphaSolid}. The volume fraction above which this occurs can be determined by:

\begin{equation}
\phi_\text{c} = \frac{4\pi\sqrt{\frac{2}{3}}}{9(1+q)^3}
\quad\text{,}
\label{osmoticPresRelation}
\end{equation}

\noindent which results in $\phi_\text{c} = 0.66$ for $q = 0.2$, indicated by the black dot in Fig. \ref{fig5}b. For $q = 0.4$, multiple overlap of depletion zones already occurs at $\phi_\text{c} = 0.42$ and the free volume fraction strongly deviates from Eq. \ref{geometricAlphaSolid}, showing the importance of taking multiple overlap of depletion layers into account for large size ratios $q$.

\subsection{Phase behaviour of HS mixtures }

\subsubsection{Highly asymmetric ($q \lesssim 0.2$)}
\begin{figure}[tb!]
	\centering
		\includegraphics[width=.48\textwidth]{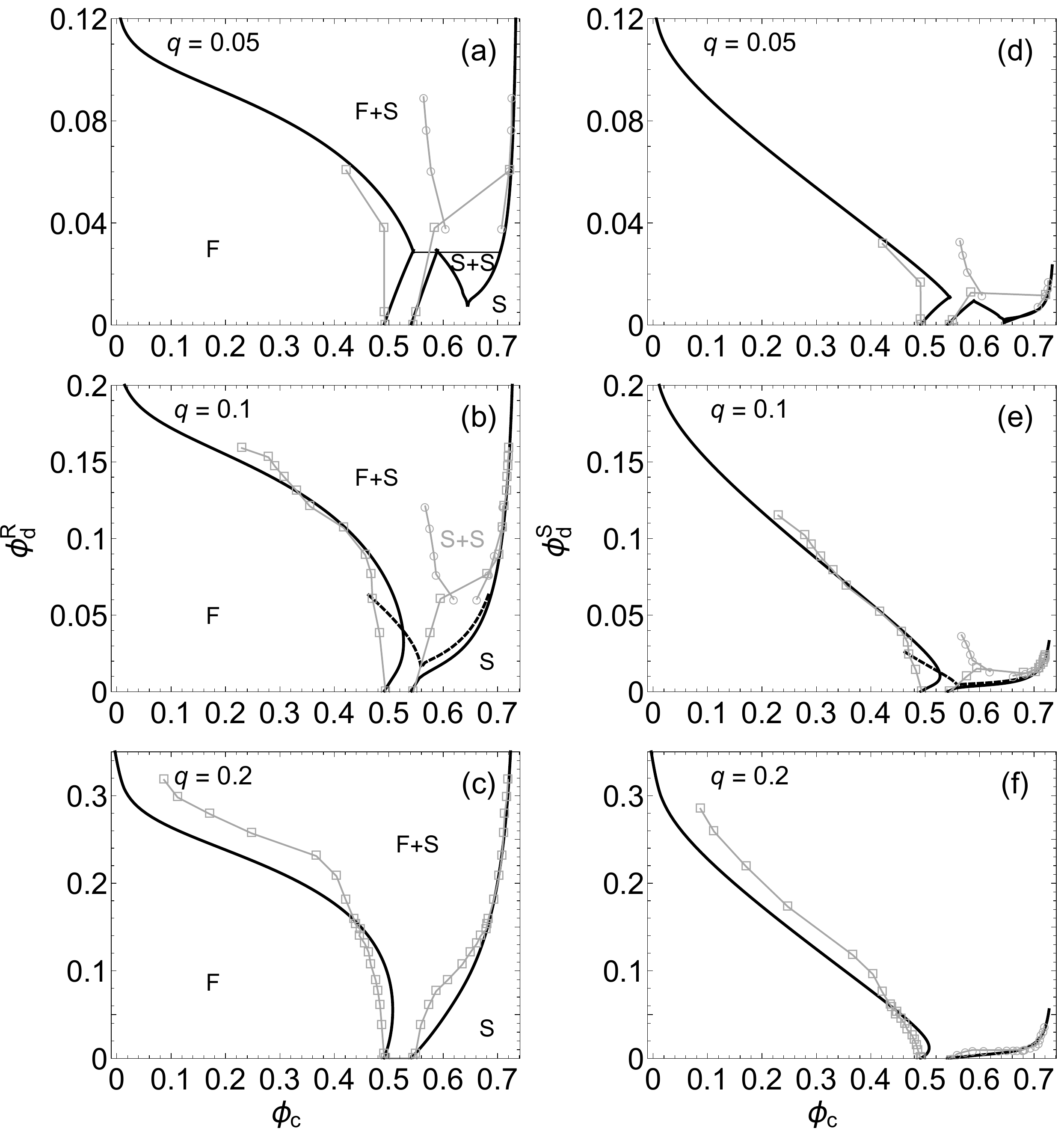}
	\caption{Phase diagrams of binary hard-sphere mixtures for $q = 0.05$ (a,d), $q = 0.1$ (b,e) and $q = 0.2$ (c,f) in the reservoir representation (a-c) and the system representation (d-f). The black curves show the coexistence lines determined with FVT, the gray data points are results of direct coexistence simulations from Dijkstra \textit{et al.} \cite{Dijkstra1999}. The open squares denote stable coexistence densities and the open circles denote metastable coexistence densities. The gray curves are a guide to the eye. The dashed black curves in (b) and (e) show the metastable isostructural solid-solid coexistence lines from FVT. For $q = 0.05$ and $q = 0.1$, Eq. \ref{alphamixs2} was used for $\alpha^\text{S}_\text{solid}$ and for $q = 0.2$ the numerical procedure as described in Sec. \ref{sssec:223} was used. Phase regions are indicated in (a-c); the fluid phase is denoted with F, the solid phase with S, the fluid-solid coexistence region with F+S and the isostructural solid-solid coexistence region with S+S.}
	\label{fig6}
\end{figure}

\noindent Phase diagrams were computed for binary hard-sphere mixtures with size ratio $q = 0.05, 0.1$ and $0.2$ using the semi-grand potential descriptions given by Eqs. \ref{endOmegaFluid} and \ref{endOmegaSolid}. The free volume fraction in the solid phase of the binary mixture is described using Eq. \ref{alphamixs2}. For $q = 0.05$ and $q = 0.1$, the free volume fraction of the one-component solid $\alpha_\text{solid}$ given by Eq. \ref{geometricAlphaSolid} is used. Multiple overlap of depletion zones is possible for $q = 0.2$ at high densities, therefore $\alpha_\text{solid}$ is determined following the numerical procedure described in Sec. \ref{sssec:223}. It is noted that the phase diagram for $q = 0.2$ determined with Eq. \ref{geometricAlphaSolid} showed no significant difference from the phase diagram calculated with the numerical $\alpha_\text{solid}$, which is most likely due to the fact that the deviations between both methods are quite small for this size ratio as shown in Fig. \ref{fig5}b. A comparison with the theoretical phase diagrams and phase coexistence data obtained from direct coexistence simulations from Dijkstra \textit{et al.} \cite{Dijkstra1999} is shown in Fig. \ref{fig6} for both the reservoir and the system representation. Metastable isostructural coexistence lines are not shown in the theoretical phase diagrams except for the solid-solid phase coexistence for $q = 0.1$.

\begin{figure}[tb!]
	\centering
		\includegraphics[width=.48\textwidth]{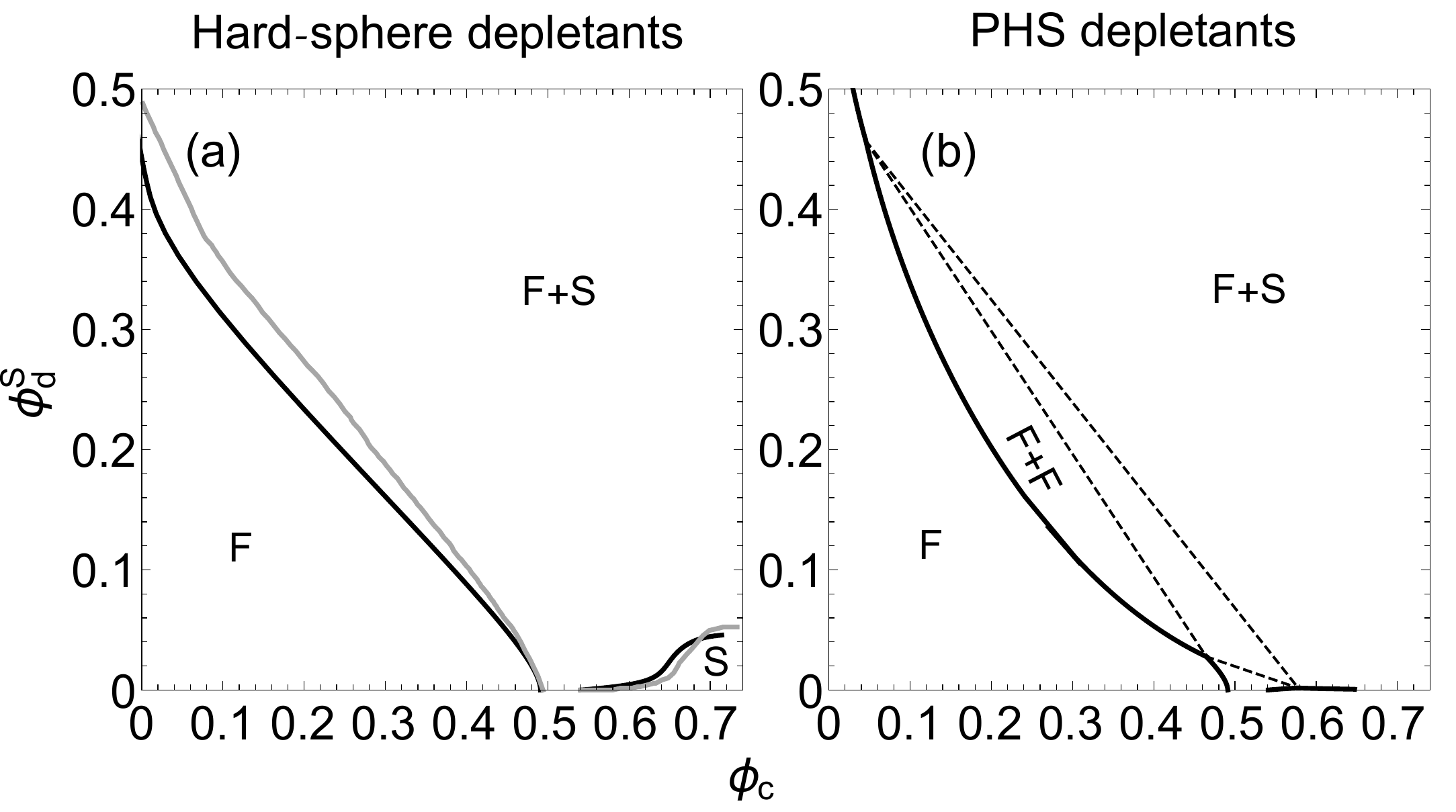}
	\caption{Phase diagram of a binary hard-sphere mixture with size ratio $q = 0.4$ in the system representation. In (a), the black curve shows the results of FVT for hard-sphere depletants with Eq. \ref{alpha04mix} for the free volume fraction in the solid phase and the gray curve shows results from Monte Carlo free energy computer simulations\cite{Filion2011}. Coexistence lines with phases where the small particles crystallize ($\phi_\text{d}^\text{S} > 0.49$) are not shown. The FVT results from the PHS approximation, with Eq. \ref{alfa04} for the free volume fraction in the solid phase, are shown in (b). The dashed lines indicate the fluid-fluid-solid triple coexistence region. the fluid phase is denoted with F, the solid phase with S, the fluid-solid coexistence region with F+S and the isostructural fluid-fluid coexistence region with F+F.}
	\label{fig7}
\end{figure}

The theoretical binodals are in qualitative agreement with the simulation results. The binodals shift to lower depletant concentrations when the size ratio $q$ becomes smaller and an isostructural solid-solid coexistence region appears at low values of $q$. The solid-solid coexistence region in the theoretical phase diagram is metastable for $q = 0.1$, whereas a small stable isostructural solid-solid coexistence region was found in simulations. The discrepancy between the solid-solid coexistence regions and the mismatch of the fluid-solid binodals at low depletant concentrations is mostly likely because the geometrical description of the free volume fraction in the solid phase becomes less accurate for low packing fractions $\phi_\text{c}$ since a perfect FCC crystal is assumed. For $q = 0.2$ there is a slight underestimation of the fluid branch of the binodal compared to the simulation data. Overall, the phase diagrams are in much better agreement with the simulation data than the original FVT for binary hard-sphere mixtures \cite{Lekkerkerker1993,Poon1994}, as can be seen for example by comparing Fig. \ref{fig6}b with Fig. \ref{fig2}. Moreover, the FVT phase diagrams presented in Fig. \ref{fig6} are very similar to phase diagrams determined with FVT using the PHS approximation, and Eq. \ref{geometricAlphaSolid} for the free volume fraction in the solid phase, which is in line with the perturbation theory predictions of Velasco \textit{et al.} \cite{Velasco1999}. The agreement of the phase diagrams obtained with the FVT presented in this paper with simulations \cite{Dijkstra1999} and previous perturbation and DFT studies \cite{Velasco1999,Roth2001} indicates that the excluded volume of the depletants is now accurately taken into account and FVT can be accurately applied to hard depletants. For future applications, FVT can be extended to study the phase behaviour of colloidal mixtures containing hard anisotropic depletants. It must be noted that anisotropic depletants have already been studied with FVT \cite{Vliegenthart1999,Oversteegen2005,Oversteegen2004a}, however the same approximations are made in these studies as the approximations in the original FVT for binary hard-sphere mixtures. It is expected that explicitly taking the excluded volume of the depletants into account in FVT has a large influence on the phase behaviour predictions for anisotropic depletants since these have a larger effective excluded volume than spherical depletant particles. 

\subsubsection{Interstitial solid solution ($q = 0.4$)}
\noindent In Fig. \ref{fig7}a, we present the phase diagram of a binary hard-sphere mixture with a size ratio of $q = 0.4$, determined using the numerical free volume fraction given by Eq. \ref{alpha04mix}. Also shown is the phase diagram resulting from Monte Carlo free energy simulations by Filion \cite{Filion2011}. Fig. \ref{fig7}b shows the FVT results from the PHS approximation, but with multiple overlap taken into account by using Eq. \ref{alfa04} for $\alpha_\text{solid}$. It is noted that we only focus on the fluid phase and the interstitial solid solution (ISS). At large depletant concentrations, $\phi_\text{d} > 0.49$, the small depletant particles form an FCC crystal which the large particles cannot enter. For these concentrations, simulations predict a coexistence between the ISS phase and the small particle FCC crystal, and a triple coexistence region where these two phases coexist with the binary fluid. As can be seen in Fig. \ref{fig7}b, the PHS approximation also predicts a fluid-fluid coexistence region, which is not found in simulations for hard-sphere depletants \cite{Filion2011}. This shows that the PHS approximation no longer accurately describes the phase behaviour of the binary mixture in contrast to the size ratios $q \lesssim 0.2$. The reason for the absence of the fluid-fluid coexistence region for hard-sphere depletants can be understood by comparing the pair potentials for PHS and hard-sphere depletants in Fig. \ref{fig1}b. Fluid-fluid coexistence requires a long range attraction and Fig. \ref{fig1}b shows that the range of the primary minimum is much smaller for a hard-sphere depletant compared to a PHS depletant. The region of fluid-fluid coexistence is not found with the FVT approach for a binary hard-sphere mixture described in this paper. The binodals of the adjusted FVT approach again compare qualitatively well with the simulation results, indicating that the proposed method of this paper can be extended beyond highly asymmetric binary hard-sphere mixtures. 

\subsection{Outlook}
\noindent It is interesting to extend our approach to binary colloidal mixtures in which interactions beyond hard core interactions are accounted for. In this paper the focus is on binary mixtures of hard spheres, while classical FVT focused on hard-sphere + PHS mixtures\cite{Lekkerkerker1992}. It may be interesting to vary the additivity of the depletants to investigate the transition from PHS (non-additive) to HS (fully additive), along the lines of Roth and Evans\cite{Roth2001a}. Also, it is useful to vary, for an asymmetric binary mixture only the degree of the interactions between the dissimilar particles as Wilding \textit{et al.}\cite{Wilding1998} did for a symmetrical mixture. A further step may be to induce a simple Baxter\cite{Baxter1968} stickiness between the particles\cite{Noro1999,Fantoni2015}. That would involve three stickiness parameters. Finally, this may be extended by investigating binary hard-core Yukawa (HCY) mixtures, where each hard-core particle species interacts with its own and other type of particles through an additional Yukawa interaction with adjustable sign, strength and range. This has been done for HCY + PHS mixtures\cite{GonzalezGarcia2016}. An important element for all these extensions is to obtain knowledge of the preferred solid structures that appear in such mixtures, for which the simulation method of Filion and Dijkstra\cite{Filion2009} may provide a useful starting point.  

\section{Conclusions}
\noindent A FVT approach that explicitly takes the excluded volume of hard-sphere depletants into account was developed. The descriptions of the free volume fractions in a highly asymmetric binary hard-sphere fluid and solid were verified by comparing the volume fraction of depletants in the system as a function of the volume fraction in the reservoir with computer simulation results. Explicitly taking the excluded volume of the depletants into account leads to a significantly better match with the simulation data than original FVT for binary hard-sphere mixtures. Moreover, the phase diagrams obtained with the FVT approach presented in this paper are in qualitative agreement with simulation results. For highly asymmetric mixtures, FVT for hard-sphere depletants and PHS depletants lead to very similar results as expected from perturbation theory.

FVT is more difficult to apply for large size ratios due to the possibility of multiple overlap of depletion zones ($q > 0.15$) and the wide variety of binary solid phases that can be formed for $q > 0.2$. However, we have shown that the phase behaviour of a binary hard-sphere mixture with a size ratio of $q = 0.4$, where a simple interstitial solid solution is formed at high densities, can be described reasonably well using FVT. Although it is not possible to develop a general FVT method for binary hard-sphere mixtures with size ratios $q > 0.2$, a similar approach as for $q = 0.4$ could in principle be followed for specific size ratios, as long as the binary solid phases that can be formed are known in advance. 

\begin{acknowledgments}
\noindent We thank professors Dirk Aarts and Pavlik Lettinga for fruitful discussions. The authors are grateful for financial support from the Dutch Ministry of Economic Affairs of the Netherlands via The Top-consortium Knowledge and Innovation (TKI) roadmap Chemistry of Advanced Materials (CHEMIE.PGT.2018.006). 
\end{acknowledgments}

\section*{Data Availability}
\noindent All data computed in this work and copies of the Mathematica codes are available upon request.

\nocite{*}


\begin{thebibliography}{63}
\bibitem[{num()}]{numalpha}
\bibinfo{note}{A unit cell can be manually created by placing spheres with
  radius $R_\text{c} + R_\text{d}$, representing the colloidal particles and
  their depletion zones, on the appropriate lattice positions. Next, the total
  region occupied by the depletion zones can be defined using the function
  RegionUnion and the total volume is determined using RegionMeasure. The free
  volume available in the unit cell can be directly determined by applying the
  RegionDifference function to the depletion zones region and the unit cell. An
  equation for $\alpha_\text{solid}$ for a specific value of $q$ is obtained by
  repeating this for different volume fractions $\phi_\text{c}$ and fitting the
  data with a polynomial.}
\end{thebibliography}
\end{document}